\pdfoutput=1
\documentclass[sn-mathphys]{sn-jnl}
\unnumbered
\usepackage{amsmath, siunitx}
\usepackage{amssymb}
\usepackage{bm}
\usepackage[utf8]{inputenc}
\usepackage[T1]{fontenc}
\begin{document}

\title[\empty]{Low disorder and high valley splitting in silicon}

\author[1]{\fnm{Davide} \sur{Degli Esposti}}
\author[1]{\fnm{Lucas E. A.} \sur{Stehouwer}}
\author[2]{\fnm{Önder} \sur{Gül}}
\author[2]{\fnm{Nodar} \sur{Samkharadze}}
\author[1]{\fnm{Corentin} \sur{Déprez}}
\author[1]{\fnm{Marcel} \sur{Meyer}}
\author[1]{\fnm{Ilja N.} \sur{Meijer}}
\author[2]{\fnm{Larysa} \sur{Tryputen}}
\author[2]{\fnm{Saurabh} \sur{Karwal}}
\author[3]{\fnm{Marc} \sur{Botifoll}}
\author[3,4]{\fnm{Jordi} \sur{Arbiol}}
\author[2]{\fnm{Sergey V.} \sur{Amitonov}}
\author[1]{\fnm{Lieven M. K.} \sur{Vandersypen}}
\author[2]{\fnm{Amir} \sur{Sammak}}
\author[1]{\fnm{Menno} \sur{Veldhorst}}
\author*[1]{\fnm{Giordano} \sur{Scappucci}}\email{g.scappucci@tudelft.nl}

\affil[1]{\orgdiv{QuTech and Kavli Institute of Nanoscience}, \orgname{Delft University of Technology}, \orgaddress{\street{PO Box 5046}, \city{Delft}, \postcode{2600 GA}, \country{The Netherlands}}}
\affil[2]{\orgdiv{QuTech and Netherlands Organisation for Applied Scientific Research (TNO)}, \orgaddress{\city{Delft}, \country{The Netherlands}}}
\affil[3]{\orgname{Catalan Institute of Nanoscience and Nanotechnology (ICN2), CSIC and BIST}, \orgaddress{\street{Campus UAB, Bellaterra}, \city{Barcelona}, \postcode{08193}, \country{Catalonia, Spain}}}
\affil[4]{\orgname{ICREA}, \orgaddress{\street{Pg. Lluis Companys 23}, \city{Barcelona}, \postcode{08020}, \country{Catalonia, Spain}}}

\abstract{The electrical characterisation of classical and quantum devices is a critical step in the development cycle of heterogeneous material stacks for semiconductor spin qubits. 
In the case of silicon, properties such as disorder and energy separation of conduction band valleys are commonly investigated individually upon modifications in selected parameters of the material stack. 
However, this reductionist approach fails to consider the interdependence between different structural and electronic properties at the danger of optimising one metric at the expense of the others.
Here, we achieve a significant improvement in both disorder and valley splitting by taking a co-design approach to the material stack. 
We demonstrate isotopically-purified, strained quantum wells with high mobility of $3.14(8)\times 10^5$ cm$^2$/Vs and low percolation density of $6.9(1)\times10^{10}$ cm$^{-2}$.
These low disorder quantum wells support quantum dots with low charge noise of $0.9(3)~\text{µeV/Hz}^{1/2}$ and large mean valley splitting energy of $0.24(7)$~meV, measured in qubit devices. 
By striking the delicate balance between disorder, charge noise, and valley splitting, these findings provide a benchmark for silicon as a host semiconductor for quantum dot qubits. We foresee the application of these heterostructures in larger, high-performance quantum processors.}

\keywords{silicon, valley splitting, mobility, charge noise, quantum dots}

\maketitle
\section{Introduction}

The development of fault-tolerant quantum computing hardware relies on significant advancements in the quality of quantum materials hosting qubits\cite{de_leon_materials_2021}. 
For spin qubits in gate-defined silicon quantum dots\cite{vandersypen_interfacing_2017}, there are currently three material-science driven requirements being pursued\cite{scappucci_crystalline_2021}. 
The first is to minimise potential fluctuations arising from static disorder in the host semiconductor, to ensure precise control of the charging energies and tunnel coupling between quantum dots, and to enable shared control in crossbar arrays\cite{li_crossbar_2018}. 
The second requirement is to reduce the presence of two-level fluctuators and other sources of dynamic disorder responsible for charge noise, which currently limits qubit performance\cite{philips_universal_2022,yoneda_quantum-dot_2018}. 
Lastly, it is crucial to maximise the energy separation between the two low-lying conduction valleys\cite{zwanenburg_silicon_2013}. 
Achieving large valley splitting energy prevents leakage outside the computational two-level Hilbert space and is essential to ensure high fidelity spin qubit initialisation, readout, control, and shuttling\cite{vandersypen_interfacing_2017,tagliaferri_impact_2018,huang_spin_2014,seidler_conveyor-mode_2022,langrock_blueprint_2023,zwerver_shuttling_2023}.
Satisfying these multiple requirements simultaneously is challenging because the constraints on material stack design and processing conditions may conflict. 
In gate-defined silicon quantum dots, single electron spins are confined either at the semiconductor-dielectric interface in metal-oxide-semiconductor (Si-MOS) stacks or in buried strained quantum wells at the hetero-epitaxial Si/SiGe interface. 
In Si-MOS, the large electric field at the interface between the semiconductor and the dielectric drives a large valley splitting energy in tightly confined quantum dots\cite{yang_spin-valley_2013}. 
However, the proximity of the dielectric interface induces significant static and dynamic disorder, affecting mobility, percolation density, and charge noise\cite{sabbagh_quantum_2019}. 
The latter can be improved through careful optimisation of the multi-layer gate stack resorting to industrial fabrication processes\cite{elsayed_low_2022}. 

In conventional Si/SiGe heterostructures, a strained Si quantum well is separated from the semiconductor-dielectric interface by an epitaxial SiGe barrier\cite{scappucci_crystalline_2021}.
The buried Si quantum well naturally ensures a quiet environment, away from the impurities at the semiconductor-dielectric interface, leading to lower disorder and charge noise compared to Si-MOS\cite{borselli_measurement_2011,degli_esposti_wafer-scale_2022,paquelet_wuetz_reducing_2023}. 
However, strain and compositional fluctuations in the SiGe strain-relaxed buffer (SRB) below the quantum well result in band-structure variations and device non-uniformity\cite{evans_nanoscale_2012}. 
Furthermore, valley splitting is limited and may vary from device to device\cite{zajac_reconfigurable_2015,shi_tunable_2011,scarlino_dressed_2017,ferdous_valley_2018,mi_high-resolution_2017,mi_circuit_2017,mi_landau-zener_2018,kawakami_electrical_2014} due to the weaker electric field compared to Si-MOS\cite{Fiesen2007VS_theory,lodari_valley_2022} and the additional in-built random alloy composition fluctuations at the strained Si-SiGe hetero-interface\cite{paquelet_wuetz_atomic_2022}, posing a challenge for device reliability and qubit operation. 

Practical strategies have been recently considered to enhance valley splitting in Si/SiGe quantum wells\cite{losert_practical_2023}, including the use of unconventional heterostructures that incorporate Ge to the interior\cite{mcjunkin_valley_2021,feng_enhanced_2022,paquelet_wuetz_atomic_2022,mcjunkin_sige_2022} or the boundary of the quantum
well\cite{zhang_genetic_2013,neyens_critical_2018,wang_origin_2022}. 
Without a co-design for high electron mobility, enhancing valley splitting, which requires breaking translation symmetry, tends to occur at the expense of a deteriorated disorder landscape, posing challenges for scaling to large qubit systems. 
Indeed, the few experimental reports\cite{borselli_measurement_2011,hollmann_large_2020,mcjunkin_sige_2022} of large valley splitting (e.g. $> 0.2$~meV) in Si/SiGe quantum dots have shown relatively low mobility ($< 6 \times 10^4$~cm$^2$/Vs) of the parent two-dimensional electron gas, thereby spoiling one major advantage of Si/SiGe over Si-MOS. 
A large valley splitting up to $0.239$~meV has been measured in quantum wells incorporating an oscillating Ge concentration\cite{mcjunkin_sige_2022}. 
However, the additional scattering from random alloy disorder yields an electron mobility of $2 - 3 \times 10^4$~cm$^2$/Vs. This mobility is significantly lower than what is obtained with conventional Si/SiGe heterostructures\cite{mi_magnetotransport_2015,paquelet_wuetz_multiplexed_2020} and is even comparable to the mobility in the best Si-MOS stacks\cite{elsayed_low_2022}. 
Instances of large valley splittings (up to $0.286\pm0.026$~meV) within a wide distribution have also been measured in $3$~nm ultra-thin quantum wells\cite{chen_detuning_2021}. 
Likewise, ultra-thin quantum wells may degrade mobility due to increased scattering from random alloy disorder as the wave function penetrates deeper into the SiGe barrier\cite{gold_barrier_1989}, potentially compounded by interface roughness as well\cite{ismail_identification_1994}.
Conversely, very high mobility of $ 6.5 \times 10^5$~cm$^2$/Vs was reported in conventional Si/SiGe heterostructures 
although the quantum dots showed rather low valley splitting in the range of $35-70~\text{µeV}$\cite{mi_circuit_2017}.

In this work, we present significant advancements in isotopically purified $^{28}$Si/SiGe heterostructures by conducting a study across multiple Hall bars and quantum dots in spin-qubit devices. 
We demonstrate simultaneous improvement in the channel static disorder, qualified by mobility and percolation density, and in the mean valley splitting while keeping respectable levels of low-frequency charge noise. 
These advancements are achieved without resorting to unconventional heterostructures. 
Instead, they result from explicitly accounting for the unavoidable broadening of Si-SiGe interfaces and optimising the quantum well thickness, while considering the design constraints imposed by the chemical composition of the SiGe buffer.
Specifically, we ensure that the quantum well thickness is chosen to maintain coherent epitaxy of the strained Si layer with the underlying SiGe buffer while also minimising the impact of disorder originating from barrier penetration effects.

\section{Results}

\subsection{Description of the heterostructures}

The $^{28}$Si/SiGe heterostructures are grown on a 100~mm Si(001) substrate by reduced-pressure chemical vapour deposition (Methods). 
From bottom to top (Fig.~1a), the heterostructure comprises a thick SiGe strained relaxed buffer (SRB) made of a step graded Si$_{1-x}$Ge$_x$ buffer layer with increasing Ge concentration followed by a SiGe layer with constant Ge concentration, a tensile-strained $^{28}$Si quantum well, and a SiGe barrier passivated by an amorphous Si-rich layer\cite{degli_esposti_wafer-scale_2022}. 
Given the in-plane random distribution of Si and Ge at the interfaces between Si and SiGe layers, the description of a realistic Si quantum well may be reduced to the one-dimensional Ge concentration profile along the growth direction\cite{paquelet_wuetz_atomic_2022,dyck_accurate_2017}. 
This is modelled by sigmoidal interfaces\cite{dyck_accurate_2017} (Methods) as in Fig~1b and is characterised by three parameters: $\rho_{\mathrm{b}}$ is the asymptotic limit value of the maximum Ge concentration in the SiGe barriers surrounding the quantum well; $4\tau$ is the interface width, which corresponds to the length over which the Ge concentration changes from 12\% to 88\% of $\rho_{\mathrm{b}}$; $w$ is the quantum well width defined as the distance between the inflection points of the two interfaces. 
Our growth protocol yields a reproducible quantum well profile with $\rho_{\mathrm{b}}=0.31(1)$\cite{xue_quantum_2022,paquelet_wuetz_atomic_2022}, $4\tau \approx 1 \textnormal{~nm}$, and $w \approx 7\textnormal{~nm}$ (see Supplementary Figs.~1 and 2). 
The quantum well thickness was chosen on purpose to fall within the range of 5~nm to 9~nm, which correspond to the thicknesses of quantum wells studied in ref.~\cite{paquelet_wuetz_reducing_2023} and used here as a benchmark. 
We expect a quantum well of about 7~nm to be thin enough to suppress strain-release defects and also increase the valley splitting compared to the results in ref~\cite{paquelet_wuetz_atomic_2022,philips_universal_2022,xue_quantum_2022,noiri_fast_2022}. 
At the same time, the quantum well was chosen to be sufficiently thick to mitigate the effect of disorder arising from penetration of the wave function into the SiGe barrier\cite{gold_barrier_1989} and possibly from the interface roughness\cite{ismail_identification_1994}. 

Figure~1d shows aberration corrected (AC) atomic resolution high angle annular dark field (HAADF) scanning transmission electron microscopy (STEM) images and superimposed intensity profiles to validate the thickness of the $^{28}$Si quantum well by counting the (002) horizontal planes as in ref.~\cite{paquelet_wuetz_reducing_2023}. 
We estimate that the quantum well is formed by $26$ atomic planes, corresponding to a thickness $w = 6.9 \pm 0.5$ nm (see Supplementary Fig.~1). 
Further electron microscopy characterisation of all quantum wells considered in this study highlights the robustness of our growth protocol (see Supplementary Fig.~2).
Images in Fig.~1d,e, acquired in HAADF (Z-contrast) and bright field (BF) STEM modes, respectively, highlight two critical characteristics of the compositionally graded SiGe layers beneath the quantum well. 
Firstly, the step-wise increase of the Ge content corresponds clearly to the varying shades of contrast in Fig.~1d. 
Secondly, strain-release defects and dislocations in Fig.~1e are confined at the multiple and sharp interfaces within the compositionally graded buffer layer, highlighting the overall crystalline quality of the SiGe SRB below the quantum well.

\subsection{Characterisation of strain distribution}

After confirming the quantum well thickness, we examine the coherence of the Si quantum well epitaxy with the underlying SiGe and quantify the in-plane strain ($\epsilon$) of the quantum well, along with the amplitude ($\Delta\epsilon$) of its fluctuations. 
Following the approach in ref.~\cite{kutsukake_origin_2004}, we employ scanning Raman spectroscopy on a heterostructure where the SiGe top barrier is intentionally omitted.
Since this configuration maximises the signal from the thin strained Si quantum well, we are able to efficiently map the shift in Si-Si vibrations originating from both the Si quantum well ($\omega_{\textnormal{Si}}$) and from the SiGe buffer layer below ($\omega_{\textnormal{SiGe}}$) (see Supplementary Fig.~3). 
Fig.~2a shows an atomic-force microscopy image of a pristine grown $^{28}$Si/SiGe heterostructure over an area of $90\times 90~\text{µm}^2$. 
The surface is characterised by a root mean square (RMS) roughness of $\approx 2.4$~nm and by the typical cross-hatch pattern arising from the misfit dislocation network within the SiGe SRB. 
The cross-hatch undulations have a characteristic wavelength of $\approx 5~\text{µm}$ estimated from the Fourier transform spectrum.

The Raman map in Fig.~2b tracks $\omega_{\textnormal{Si}}$ over an area of $40\times 40~ \text{µm}^2$. This area is sufficiently large to identify fluctuations due to the cross-hatch pattern in Fig.~2a, with regions featuring higher and lower Raman shifts around a mean value of $\overline{\omega}_{\textnormal{Si}} = 510.4(2)$~cm$^{-1}$. In Fig.~2c, we investigate the relationship between the Raman shifts from the quantum well $\omega_{\textnormal{Si}}$ and from the SiGe buffer $\omega_{\textnormal{SiGe}}$. We find a strong linear correlation with a slope $\Delta\omega_{\textnormal{Si}}/\Delta\omega_{\textnormal{SiGe}} = 1.01(2)$, suggesting that the distribution of the Raman shift in the Si quantum well is mainly driven by strain fluctuations in the SiGe SRB, rather than compositional fluctuation\cite{kutsukake_origin_2004}. 

We calculate the strain in the quantum well using the equation $\epsilon = (\omega_{\textnormal{Si}} - \omega_\textnormal{0}) / b_\textnormal{Si}$, where $\omega_\textnormal{0} = 520.7$~cm$^{-1}$ is the Raman shift for bulk, relaxed Si and $b_\textnormal{Si} = 784(4)$ cm$^{-1}$ is the Raman phonon strain shift coefficient of strained silicon on similar SiGe SRBs\cite{wong_determination_2005}. From $\overline{\omega}_{\textnormal{Si}}$, we estimate the mean value of the in-plane strain for the quantum well $\overline{\epsilon} = 1.31(3)$~\%. This value is qualitatively comparable to the expected value of $\approx 1.19(4)$~\% from the lattice mismatch between Si and the Si$_{0.69}$Ge$_{0.31}$ SRB (see Supplementary Note 2). A more quantitative comparison would require a direct measurement of $b_\textnormal{Si}$ on our heterostructures based upon high-resolution X-ray diffraction analysis across multiple samples with varying strain conditions. Figure~2d shows the normalised distribution of strain fluctuations percentage around the mean value $\Delta \epsilon / \overline{\epsilon} = (\epsilon -\overline{\epsilon})/\overline{\epsilon}$. 
The data follows a normal distribution (black line) characterised by a standard deviation of $3.0(1)$ \%, comparable with similar measurements in strained Ge/SiGe heterostructures\cite{stehouwer_germanium_2023}. 
Given the significant correlation between Raman shifts in the quantum well and the SiGe buffer, alongside the measured strain levels exhibiting a narrow bandwidth of fluctuations, we argue that, with our growth conditions, the Si quantum well is uniformly and coherently grown on the underlying SiGe buffer. As a consequence, we expect strain-release defects in the quantum well to be very limited, if present at all.

\subsection{Electrical characterisation of heterostructure field effect transistors}

We evaluate the influence of the design choice of a 7~nm thick quantum well on the scattering properties of the 2D electron gas (2DEG) through wafer-scale electrical transport measurements. 
The measurements are performed on Hall-bar-shaped Heterostructure Field-Effect Transistors (H-FETs) operated in accumulation mode (Methods). 
Multiple H-FETs across the wafer are measured within the same cool-down at a temperature of 1.7~K using refrigerators equipped with cryo-multiplexers\cite{paquelet_wuetz_multiplexed_2020}.
Figure~3a,b show the mean mobility-density and conductivity-density curves in the low-density regime relevant for quantum dots. These curves are obtained by averaging the mobility-density curves from ten H-FETs fabricated from the same wafer (solid line), and the different shadings represent the intervals corresponding to one, two, and three standard deviations. The distribution of mobility and conductivity is narrow, with a variance lower than 5\% over the entire density range. 
Furthermore, we observe similar performance from H-FETs fabricated on a nominally identical heterostructure grown subsequently (see Supplementary Fig.~5), indicating the robustness of both our heterostructure growth and H-FET fabrication process. At low densities, the mobility increases steeply due to the increasing screening of scattering from remote impurities at the semiconductor-dielectric interface. This is confirmed by the large power law exponent $\alpha = 2.7$ obtained by fitting the mean mobility-density curve to the relationship $\mu \propto n^\alpha$ in the low-density regime\cite{laroche_scattering_2015}. At high density, the mobility keeps increasing, albeit with a much smaller power law exponent $\alpha = 0.3$. This indicates that scattering from nearby background impurities, likely oxygen within the quantum well\cite{mi_magnetotransport_2015}, and potentially interface roughness\cite{huang_understanding_2023} become the limiting mechanisms for transport in the 2DEG. 

From the curves in Fig.~3a,b, we obtain the distributions of mobility $\mu$ measured at high density ($n = 6\times10^{11}$~cm$^{-2}$) and of the percolation density $n_\textnormal{p}$, extracted by fitting (black line) to percolation theory\cite{tracy_observation_2009}. In Fig.~3c,d, we benchmark these metrics for the 6.9 nm thick quantum well against the distributions obtained previously\cite{paquelet_wuetz_reducing_2023} for a quantum well thickness of $5.3(5)$ nm and $9.0(5)$ nm. 
The $6.9$~nm quantum well performs the best, with a mean mobility at high densities of $\mu = 3.14(8) \times 10^5$~cm$^2$/Vs and a percolation density of $n_\textnormal{p} = 6.9(1) \times 10^{10}$~cm$^{-2}$.

The distributions show two noteworthy features: a 50~\% increase in mobility between the $5.3$ nm and $6.9$ nm quantum well and a three-fold reduction in the variance of the distribution between the $9.0$ nm quantum well and the remaining two. We attribute the mobility increase to reduced scattering from alloy disorder, as the wave function delocalizes further into the quantum well rather than penetrating into the barrier\cite{gold_barrier_1989}. 
We attribute the large spread in transport properties of the widest quantum well to some degree of strain relaxation and associated defects\cite{paquelet_wuetz_reducing_2023}. This explanation is further supported by comparative measurements of Raman shift correlation (see Supplementary Fig.~4) and highlights the sensitivity of the transport properties and their distributions to strain relaxation in the quantum well.

\subsection{Charge noise measurements in quantum dots}

Moving on to quantum dot characterisation, we focus on the measurement of low-frequency charge noise using complete spin qubit devices cooled at the base temperature of a dilution refrigerator (Methods). 
The device design is identical to the one in refs.~\cite{unseld_2d_2023,meyer_electrical_2023} and features overlapping gates for electrostatic confinement and micromagnets for coherent driving. We tune the sensing dot in the single electron regime, measure time traces of the source-drain current $I_{\textnormal{SD}}$ on a flank of a Coulomb peak, and repeat for several peaks before the onset of a background current. From the time-dependent $I_{\textnormal{SD}}$ we obtain the current noise power spectral density $S_{\textnormal{I}}$ and convert to charge noise power spectral density $S_{\epsilon}$ using the measured lever arm and slope of each Coulomb peak (Methods). We confirm that chemical potential fluctuations are the dominant contributions to the noise traces by measuring the noise in the Coulomb blockade and on top of a Coulomb peak (see Supplementary Fig.~6)\cite{connors_low-frequency_2019}. The latter measurement also excludes that the noise traces have any relation to the change of noise floor of the current amplifier\cite{lodari_low_2021}.

Figure~4a shows a representative noise spectrum. We observe an approximate $1/f$ trend at low frequency, suggesting the presence of an ensemble of two-level fluctuators (TLFs) with a wide range of activation energies\cite{dutta_energy_1979,ahn_microscopic_2021}. Notably, a kink appears at a specific frequency, which is attributed to the additional contribution in the power spectral density of a single TLF near the sensor\cite{connors_low-frequency_2019,elsayed_low_2022}. 
We fit this spectrum to a function which is the sum of a power law and a Lorentzian of the form $\frac{A}{f^{\alpha}} + \frac{B}{f/f_0 + 1}$, where A, B, $\alpha$, and $f_0$ are fitting parameters. 
We extract $f_0 = 10.38(3)~\text{Hz}$, $\alpha = 1.66(2)$, and the power spectral density at 1~Hz $S_{\epsilon} (\textnormal{1 Hz}) = 0.60(5)~\text{µeV/Hz}^{1/2}$. We repeat the analysis on a set of 17 noise spectra obtained from measurements of two separate devices (Supplementary Figs.~7 and 8). We do not observe a clear monotonic dependence of the noise spectra on the increasing electron occupancy in the quantum dots, in agreement with the measurement in ref.~\cite{paquelet_wuetz_reducing_2023} for devices with a similar semiconductor-dielectric interface and a thinner ($w=5.3$~nm) quantum well. 

In Fig.~4b, we evaluate the noise power spectral density at 1~Hz $S_{\epsilon}^{1/2} (\textnormal{1 Hz})$ to compare the performance of the $6.9(5)$~nm and the $5.3(5)$~nm quantum well. In addition to the different thickness of the quantum well, the devices on the $5.3$~nm quantum well are defined by a single-layer of gates, whilst the devices on the $6.9$~nm quantum well are complete qubit devices featuring a three-layers of overlapping gates, additional dielectric films in between, and micromagnets.
The noise power spectral density in the multi-layer devices (purple) and single-layer devices (green) are similar, with $\lvert S_{\epsilon}\rvert = 0.9(3))~\text{µeV/Hz}^{1/2}$ and $\lvert S_{\epsilon}\rvert = 0.9(9)~\text{µeV/Hz}^{1/2}$, respectively. 
Because both narrow quantum wells are fully strained, we expect the two heterostructures to contribute similarly to the electrostatic noise. Therefore, our measurements suggest that using multiple metallic gates, dielectric layers, and micromagnets does not degrade the noise performance in our devices. Our observations are consistent with previous measurements in Si/SiGe heterostructures at base temperature when impurities in the dielectric likely freeze out\cite{undseth_hotter_2023,connors_low-frequency_2019}. We attribute this robustness to the distinctive characteristics of Si/SiGe heterostructures, where the active region of the device resides within a buried quantum well, well separated from the gate stack, unlike Si-MOS. We speculate that the metallic layers in the gate stack, positioned between the quantum well and the micromagnets, may shield the effects of additional impurities and traps in the topmost layers. 

\subsection{Valley splitting measurements in quantum dots}

To complete the quantum dot characterisation, we measure the two-electron singlet-triplet splitting $E_{\textnormal{ST}}$ in quantum dot arrays as in the six spin qubit devices described in ref.~\cite{philips_universal_2022} by mapping the 1e $\rightarrow$ 2e transition as a function of the parallel magnetic field ($B$). $E_{\textnormal{ST}}$ is a reliable estimate of the valley splitting energy $E_{\textnormal{V}}$ in strongly confined quantum dots\cite{zajac_reconfigurable_2015,ercan_strong_2021,dodson_how_2022,paquelet_wuetz_atomic_2022} and is the relevant energy scale for spin-to-charge conversion readout with Pauli spin-blockade\cite{borselli_pauli_2011,philips_universal_2022}.

Figure~5a shows a typical magnetospectroscopy map with a superimposed thin line highlighting the 1e $\rightarrow$ 2e transition at a given magnetic field (Methods). The thick line is a fit of the transition to the theoretical model\cite{paquelet_wuetz_atomic_2022,dodson_how_2022}, allowing us to estimate the singlet-triplet splitting $E_{\textnormal{ST}} = g \mu_\textnormal{B} B_{\textnormal{ST}}$. 
Here, $g = 2$ is the electron gyromagnetic ratio, $\mu_{\textnormal{B}}$ is the Bohr magneton, and $B_{\textnormal{ST}}$ corresponds to the magnetic field at which the energy of the 1e $\rightarrow$ 2e transition starts to decrease, signalling the transition from the singlet state $S_0$ to the triplet state ($T_{-}$) as the new ground state of the two-electron system. 
For this specific quantum dot, we find $B_{\textnormal{ST}} = 1.77(2)$~T, corresponding to $E_{\textnormal{V}} = 0.205(2)$~meV.

Figure 5b compares the valley splitting of spin qubit devices on the $6.9$~nm quantum well (purple, see Supplementary Fig.~9) and on the $9.0$~nm quantum well (blue) from ref.~\cite{philips_universal_2022}. While the dots in all devices measured have the same nominal design and share the same fabrication process (Methods), the heterostructures further differ in the passivation of the SiGe top barrier. The heterostructure with the $6.9$~nm well is passivated by an amorphous self-terminating Si-rich layer, while the $9.0$~nm well has a conventional epitaxial Si cap\cite{degli_esposti_wafer-scale_2022} (Methods). Passivation by a self-terminating Si-rich layer yields a more uniform and less noisy semiconductor-dielectric interface, which in turn promotes higher electric fields at the Si/SiGe interface\cite{degli_esposti_wafer-scale_2022,paquelet_wuetz_reducing_2023}. We observe a statistically significant 60\% increase in the mean valley splitting in the $6.9$~nm quantum well with an amorphous Si-rich termination, featuring a mean value of $\overline{E_{\textnormal{V}}} = 0.24 \pm 0.07$~meV (see Supplementary Note 5). Furthermore, the distribution of valley splitting in devices with the wider quantum well shows instances of low values (e.g., $E_{\textnormal{V}} < 0.1 $~meV), as predicted by prevailing theory\cite{losert_practical_2023}. In contrast, these instances are absent (although still predicted) in the measured devices with the narrower quantum well.

While we cannot pinpoint a single mechanism responsible for the increase in the mean value of valley splitting, we speculate that multiple factors contribute to this observed improvement. The tighter vertical confinement within the narrower quantum well\cite{chen_detuning_2021}, coupled with the relatively wide quantum well interface width, increases the overlap of the electron wavefunction with Ge atoms in the barrier. 
This amplifies the effect of random alloy disorder, which is known to increase valley splitting\cite{paquelet_wuetz_atomic_2022,losert_practical_2023}.
Similarly, the improved semiconductor-dielectric facilitates tighter lateral and vertical confinement of the dots, which leads to a stronger electric field, contributing to drive the valley splitting\cite{Fiesen2007VS_theory,lodari_valley_2022}. Furthermore, the near-absence (or at most very limited density) of strain-release defects in the thin quantum well ensures a smoother potential landscape, promoting improved electrostatic control and confinement of the dot. 
Additionally, we suggest that a larger amount of experimental data points is required to comprehensively explore the distribution of valley splitting in the 6.9~nm quantum well. Mapping of valley-splitting by spin-coherent electron shuttling\cite{volmer_mapping_2023}, for example, could enable a meaningful comparison with existing theory\cite{losert_practical_2023} and help determine whether the absence of low instances of valley splitting results from undersampling the distribution or is influenced by some other underlying factor.

\subsection{Discussion}

In summary, we developed strained $^{28}$Si/SiGe heterostructures providing a benchmark for silicon as a host semiconductor for gate-defined quantum dot spin qubits. Our growth protocol yields reproducible heterostructures that feature a 6.9 nm thick $^{28}$Si quantum well, surrounded by SiGe with a Ge concentration of 0.31 and an interface width of about 1~nm. These quantum wells are narrow enough to be fully strained and maintain coherence with the underlying substrate, displaying reasonable strain fluctuations. Yet, the quantum wells are sufficiently wide to mitigate the effects of penetration of the wave function into the barrier. Coupled with a high quality semiconductor-dielectric interface, these $^{28}$Si/SiGe heterostructures strike the delicate balance between disorder, charge noise, and valley splitting. We comprehensively probe these properties with statistical significance using classical and quantum devices. 
Compared to our control heterostructures supporting qubits, we demonstrate a remarkable $50$ \% increase in mean mobility alongside a $10$ \% decrease in percolation density while preserving a tight distribution of these transport properties. Our characterisation of low-frequency charge noise in quantum dot qubit devices consistently reveals low charge noise levels, featuring a mean value of power spectral density of $0.9(3)~\text{µeV/Hz}^{1/2}$ at 1~Hz. These heterostructures support consistently large valley splitting with a mean value of $0.24(7)$ meV. This is a significant advancement considering that instances of similarly large valley splitting were obtained previously on heterostructures with about one order of magnitude less mobility\cite{borselli_measurement_2011,hollmann_large_2020,mcjunkin_sige_2022}. We envisage that fine-tuning the distance between the quantum well and the semiconductor-dielectric interface, as well as the Ge concentration in the SiGe alloy, could offer avenues to further increase performance. Our findings highlight the significance of embracing a co-design approach to drive innovation in material stacks for quantum computing. As quantum processors mature in complexity, additional metrics characterising the heterostructures will likely need to be considered to optimise the design parameters and to fully leverage the advantages of the Si/SiGe platform for spin qubits.

\section{Methods}
\textbf{Si/SiGe heterostructure growth.}
The $^{28}$Si/SiGe heterostructures are grown on a 100-mm n-type Si(001) substrate using an Epsilon 2000 (ASMI) reduced-pressure chemical vapour deposition reactor. 
The reactor is equipped with a $^{28}$SiH$_4$ gas cylinder (1\% dilution in H$_2$) for the growth of isotopically enriched $^{28}$Si with 800 ppm of residuals of other isotopes\cite{sabbagh_quantum_2019}. 
Starting from the Si substrate, the layer sequence of all heterostructures comprises a step-graded Si$_{(1-x)}$Ge$_x$ layer with a final Ge concentration of $x = 0.31$ achieved in four grading steps ($x = 0.07$, $0.14$, $0.21$, and $0.31$), followed by a Si$_{0.69}$Ge$_{0.31}$ strain-relaxed buffer (SRB). 
The step-graded buffer and the SRB are $\approx 3~\text{µm}$ and $\approx 2.4~\text{µm}$ thick, respectively. 
We grow the SRB at 625~\textcelsius, followed by a growth interruption and the quantum well growth at 750~\textcelsius\cite{paquelet_wuetz_atomic_2022}. 
The various heterostructures compared in Fig.~3 of the main text differ in the thickness of the Si quantum well, which are $9.0(5)$ nm, $6.9(5)$ nm, and $5.3(5)$ nm. 
We change the thickness of the quantum well by only acting on the quantum well growth time and leaving all the other conditions unaltered. This yields heterostructures with similar interface widths (see Supplementary Figs. 1 and 2 and analysis in ref.~\cite{paquelet_wuetz_atomic_2022}).
On top of the Si quantum well, the heterostructure is terminated with a 30 nm thick SiGe spacer, grown using the same conditions as the virtual substrate. 
The surface of the SiGe spacer is passivated with DCS at 500~\textcelsius before exposure to air \cite{degli_esposti_wafer-scale_2022}.
We confirm the Ge concentration in the spacer and virtual substrate via secondary ions mass spectrometry (similar to Fig.S13 from ref.~\cite{paquelet_wuetz_atomic_2022}) and quantitative electron energy loss spectroscopy. 

\textbf{Raman spectroscopy.} The two-dimensional Raman mapping follows a similar approach as in ref.~\cite{kutsukake_origin_2004}. We perform the measurements on heterostructures where we stop the growth after the quantum well and do not grow the SiGe spacer. This maximises the Raman signal coming from the Si quantum well. The measurements were performed with a LabRam HR Evolution spectrometer from Horiba-J.Y. at the backscattering geometry using an Olympus microscope (objective x100 with a $1~\text{µm}$ lateral resolution). We use a violet laser ($\lambda = 405$ nm) and an 1800 gr/mm grating to achieve the highest spectral resolution. We focus the laser spot to have a spatial dimension of $\approx 1~\text{µm}$. Given the laser wavelength, we expect to probe the Si quantum well and the SiGe SRB below (which has a uniform composition of Ge). We calibrate the Raman shift using a stress-free single crystal Si substrate with a Raman peak position at $\omega_\textnormal{0} = 520.7$ cm$^{-1}$. We use this value as a reference for the calculation of the strain of the Si quantum well.

\textbf{Device fabrication.}
The fabrication process for H-FETs involves reactive ion etching of mesa-trench and markers; selective P-ion implantation and activation by rapid thermal annealing at 700~$^{\circ}$C; atomic layer deposition (ALD) of a 10-nm-thick Al$_2$O$_3$ gate oxide; sputtering of Al gate; selective chemical etching of the dielectric with BOE (7:1) followed by electron beam evaporation of Ti:Pt to create ohmic contacts. All patterning is done by optical lithography on a four-inch wafer scale. Single and multi-layer quantum dot devices are fabricated on wafer coupons from the same H-FET fabrication run and share the process steps listed above. Single-layer quantum devices feature all the gates in a single evaporation of Ti:Pd (3:17~nm), followed by the deposition via ALD of a 5 nm thick AlOx layer and consequent evaporation of a global top screening gate of Ti:Pd (3:27~nm). Multi-layer quantum dot devices feature three overlapping gate metallizations with increasing thickness of Ti:Pd (3:17~nm, 3:27~nm, 3:37~nm), each isolated by a 5 nm thick AlOx dielectric. Finally, a last AlOx layer of 5 nm separates the gate stack from the micro-magnets (Ti:Co, 5:200 nm). All patterning in quantum dot devices is done via electron beam lithography. 

\textbf{H-FETs electrical characterisation.}
Hall-bar heterostructure field effect transistor (H-FET) measurements are performed in an attoDRY2100 dry refrigerator equipped with cryo-multiplexer\cite{paquelet_wuetz_multiplexed_2020} at a base temperature of 1.7~K\cite{degli_esposti_wafer-scale_2022}. We operate the device in accumulation mode using a gate electrode to apply a positive DC voltage ($V_G$) to the quantum well. We apply a source-drain bias of $100~\text{µV}$ and use standard four-probe lock-in technique to measure the source-drain current $I_\textnormal{SD}$, the longitudinal voltage $V_{xx}$, and the transverse Hall voltage $V_{xy}$ as a function of $V_G$ and perpendicular magnetic field $B$. From here, we calculate the longitudinal resistivity $\rho_{xx}$ and transverse Hall resistivity $\rho_{xy}$. The Hall electron density $n$ is obtained from the linear relationship $\rho_{xy}=B/en$ at low magnetic fields. The electron mobility $\mu$ is extracted as $\sigma_{xx}=ne\mu$, where $e$ is the electron charge. The percolation density $n_p$ is extracted by fitting the longitudinal conductivity $\sigma_{xx}$ to the relation $\sigma_{xx} \propto (n - n_p)^{1.31}$\cite{tracy_observation_2009}. We invert the resistivity tensor to calculate the longitudinal ($\sigma_{xx}$) and perpendicular ($\sigma_{xy}$) conductivity.

\textbf{Low-frequency charge noise.}
We perform low-frequency charge noise measurements in a Bluefors LD400 dilution refrigerator with a base temperature of $T_\textnormal{{MC}} \approx 20$~mK. We use devices lithographically identical to those described in ref.~\cite{unseld_2d_2023}. We tune the sensing dot of the devices in the Coulomb blockade regime and use it as a single electron transistor (SET). We apply a fixed source-drain excitation to the two reservoirs connected to the SET and record the current $I_\textnormal{{SD}}$ as a function of time using a sampling rate of 1 kHz for 600 s. We measure $I_\textnormal{{SD}}$ on the flank of each Coulomb peak where $\lvert dI_\textnormal{{SD}}/dV_\textnormal{P}\rvert$ is the largest, and therefore, the SET is the most sensitive to fluctuations. We check that chemical potential fluctuations are the dominant contributions to the noise traces by measuring the noise in blockade and on top of a coulomb peak (see Supplementary Fig.6)\cite{connors_low-frequency_2019}. The latter also excludes that the noise traces have any relation to the change of noise floor of the current amplifier\cite{lodari_low_2021}. We divide the time traces into ten segments of equal length and use the Fourier transform to convert the traces in the frequency domain. We average the ten different spectral densities to obtain the final current noise spectrum in a range centred to 1 Hz between 25 mHz and 40 Hz to avoid a strong interference around 50 Hz coming from the setup. 
We convert the current noise spectrum ($S_I$) in a charge noise spectrum ($S_{\epsilon}$) using the formula \cite{paquelet_wuetz_reducing_2023,connors_low-frequency_2019}:
\begin{equation}
  S_{\epsilon} = \frac{a^2 S_I}{\lvert dI/dV_\textnormal{P}\rvert^2}
\end{equation}
where $a$ is the lever arm and $\lvert dI/dV_\textnormal{P}\rvert$ is the slope of the specific Coulomb peak selected to acquire the time trace. We calculate the lever arm from the slopes of the Coulomb diamonds as $a = \lvert \frac{m_\textnormal{S} m_\textnormal{D}}{m_\textnormal{S} - m_\textnormal{D}}\rvert$, where $m_\textnormal{S}$ and $m_\textnormal{D}$ are the slopes to source and to drain, and we estimate $\lvert dI/dV_\textnormal{P}\rvert$ from the numerical derivative of the Coulomb peak. We perform this analysis for every Coulomb peak and use the specific values of the lever arm and slope to calculate the charge noise spectrum.

\textbf{Valley splitting.}
 We perform magnetospectroscopy experiments in quantum dot devices cooled in a dilution refrigerator with a base temperature of $T_\textnormal{{MC}} \approx 10$~mK. We use devices lithographically similar to those described in ref.~\cite{philips_universal_2022}. We tune the quantum dots in the single-electron regime to isolate the 1e $\rightarrow$ 2e transition. We start the magnetospectroscopy measurement from the quantum dot closest to the sensing dot and use the remaining dots as an electron reservoir. We use the impedance of a nearby sensing dot to monitor the charge state of every quantum dot. The impedance of the sensing dot is measured using RF reflectometry. The signal is measured by monitoring the reflected amplitude of the rf readout signal through a nearby charge sensor. We use the amplitude (Device 1) and the Y component (Device 2) of the reflected signal to map the 1e $\rightarrow$ 2e transition. We fit the 1e $\rightarrow$ 2e transition as a function of the magnetic field with the relation\cite{paquelet_wuetz_atomic_2022,dodson_how_2022}:
\begin{equation}
    V_\textnormal{P} = \frac{1}{\alpha \beta_\textnormal{e}}\ln \frac{e^{\frac{1}{2} k B + \beta_\textnormal{e} E_\textnormal{{ST}}} (e^{kB} + 1) }
    {e^{kB} + e^{2kB} + e^{kB+\beta_\textnormal{e} E_\textnormal{{ST}}} + 1}
\end{equation}
where $\alpha$ is the lever arm, $V_\textnormal{P}$ is the plunger gate voltage, $E_\textnormal{{ST}}$ is the single-triplet energy splitting, $k = g \mu_\textnormal{B} \beta_e$, $\beta_e = 1/k_\textnormal{B} T_\textnormal{e}$, $g = 2$ is the $g$-factor in silicon, $\mu_\textnormal{B}$ is the Bohr magneton, $B$ is the magnetic field, $k_\textnormal{B}$ is Boltzmann's constant, and $T_\textnormal{e}$ is the electron temperature. $E_\textnormal{{ST}}$ is linked to the position of the kink ($B_\textnormal{{ST}}$) in the magnetospectroscopy traces by the relation $E_\textnormal{{ST}} = g \mu_\textnormal{B} B_\textnormal{{ST}}$.

\textbf{(Scanning) Transmission Electron Microscopy.}
For structural characterization with (S)TEM, we prepared lamella cross-sections of the quantum well heterostructures using a Focused Ion Beam (Helios 600 dual beam microscope). HR-TEM micrographs were acquired in a TECNAI F20 microscope operated at 200 kV. Atomically resolved HAADF-STEM data was obtained in a probe-corrected TITAN microscope operated at 300 kV. EELS mapping was carried out in a TECNAI F20 microscope operated at 200 kV with approximately 2 eV energy resolution and 1 eV energy dispersion. Principal Component Analysis (PCA) was applied to the spectrum images to enhance the signal-to-noise ratio.

\section{Data Availability}
\noindent All data included in this work are available from the
4TU.ResearchData international data repository at https://doi.org/10.4121/a4e3765b-9e32-492b-96fe-a9b760baef48.v1.

\section{Code Availability}
\noindent All the code used to derive the figures and analyse the data is included  with the data and is available at
4TU.ResearchData international data repository at https://doi.org/10.4121/a4e3765b-9e32-492b-96fe-a9b760baef48.v1.

\section{Acknowledgements}
\noindent We acknowledge helpful discussions with Y. Huang, G. Capellini, G. Isella, D. J. Paul, M. Mehmandoost, the Scappucci group, the Vandersypen group and the Veldhorst group at TU Delft. We thank M. Friesen for valuable comments on the manuscript. We thank V. Pajcini for the Raman spectroscopy mapping and the useful discussion.
This research was supported by the European Union's Horizon 2020 research and innovation program under the Grant Agreement No. 951852 (QLSI project) and in part by the Army Research Office (Grant No. W911NF-17-1-0274). The views and conclusions contained in this document are those of the authors and should not be interpreted as representing the official policies, either expressed or implied, of the Army Research Office (ARO), or the U.S. Government. The U.S. Government is authorised to reproduce and distribute reprints for Government purposes, notwithstanding any copyright notation herein. ICN2 acknowledges funding from Generalitat de Catalunya 2021SGR00457. ICN2 is supported by the Severo Ochoa program from Spanish MCIN/AEI (Grant No.: CEX2021-001214-S) and is funded by the CERCA Programme / Generalitat de Catalunya and ERDF funds from EU. Part of the present work has been performed in the framework of Universitat Autònoma de Barcelona Materials Science PhD program. Authors acknowledge the use of instrumentation as well as the technical advice provided by the National Facility ELECMI ICTS, node "Laboratorio de Microscopias Avanzadas" at University of Zaragoza. M.B. acknowledges support from SUR Generalitat de Catalunya and the EU Social Fund; project ref. 2020 FI 00103. We acknowledge support from CSIC Interdisciplinary Thematic Platform (PTI+) on Quantum Technologies (PTI-QTEP+).

\section{Authors Contributions Statement}
\noindent A.S. grew and designed the $^{28}$Si/SiGe heterostructures with D.D.E., L.E.A.S., and G.S.. A.S. and D.D.E. fabricated heterostructure field effect transistors measured by D.D.E. and L.E.A.S.. M.B. and J.A. performed TEM characterisation. S.V.A., L.T., and S.K. fabricated quantum dot devices under the supervision of L.M.K.V.. I.L.M., C.D., and M.M. measured the charge noise data under the supervision of M.V.. O.G. and N.S. measured valley splitting. D.D.E. analysed the data. G.S. conceived and supervised the project. D.D.E. and G.S. wrote the manuscript with input from all authors.

\section{Competing Interests Statement}
The authors declare no competing interests.




\newpage

\begin{figure}[t]
	\includegraphics[width=86mm]{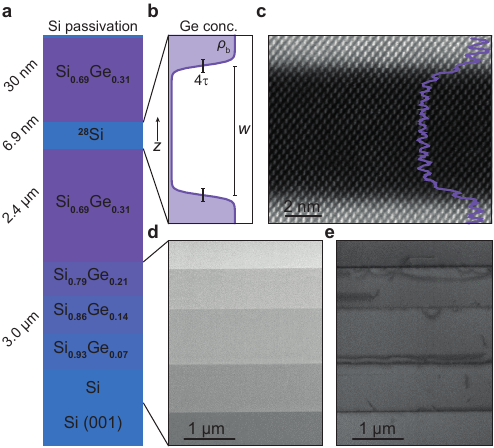}%
\caption{\textbf{Semiconductor material stack} \textbf{a} Schematic illustration of the $^{28}$Si/SiGe heterostructure. $z$ indicates the heterostructure growth direction. \textbf{b} Schematic Ge concentration profile defining a realistic Si quantum well, characterised by the final Ge concentration ($\rho_b$) in the SiGe barriers, the interface sharpness ($4\tau$), and quantum well width ($w$). 
\textbf{c} Atomic resolution high angle annular dark field (HAADF) (Z-contrast) scanning transmission electron microscopy (STEM) image of the $^{28}$Si quantum well with superimposed intensity profile used to count the number of crystallographic planes in the (002) direction forming the quantum well. \textbf{d}, \textbf{e} STEM images of the step-graded SiGe buffer layer below the quantum well acquired in HAADF (Z-contrast) and bright field mode, respectively.
}
\end{figure}

\newpage

\begin{figure}[t]
	\includegraphics[width=86mm]{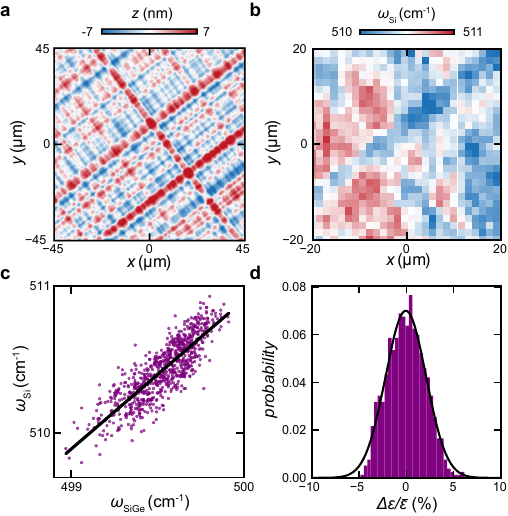}%

\caption{\textbf{Strain fluctuations measurements} \textbf{a} Atomic force microscopy image of the $^{28}$Si/SiGe heterostructure taken with an alignment of about 45 degrees to the 110 crystallographic axis. \textbf{b} Raman shift map of the Si-Si vibration $\omega_\textnormal{{Si}}$ from a strained Si quantum well with a thickness of 6.9(5)~nm. The map was taken with an alignment of about 45 degrees to the 110 crystallographic axis. \textbf{c} Cross-correlation between $\omega_\textnormal{{Si}}$ and the Si-Si vibration from the SiGe buffer ($\omega_\textnormal{{SiGe}}$) obtained by analysing Raman spectra over the same area mapped in \textbf{b} and linear fit (black line). \textbf{d} Relative in-plane strain distribution percentage of the Si quantum well $\Delta\epsilon/ \overline{\epsilon}$, where $\overline{\epsilon} = 1.31(3)$ \% is the mean value of strain in the Si quantum well. The solid line is a fit to a normal distribution.}
\end{figure}

\newpage

\begin{figure}[t]
	\includegraphics[width=86mm]{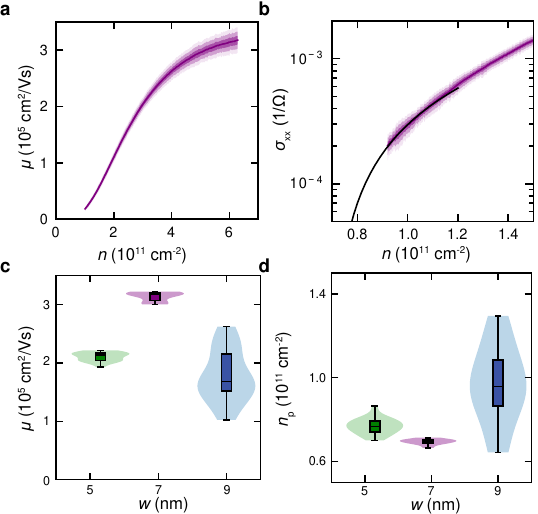}%
\caption{\textbf{Electrical transport measurements.} \textbf{a} Mean mobility $\mu$ as a function of density $n$ at $T = 1.7$ K obtained by averaging measurements from $10$ H-FETs fabricated on the heterostructure with a $6.9(5)$~nm quantum well. 
The shaded region represents one, two, and three standard deviations of $\mu$ at a fixed $n$. Data from this heterostructure are color-coded in purple in all subsequent figures.
\textbf{b} Mean conductivity $\sigma_{xx}$ as a function of $n$ and fit to the percolation theory\cite{tracy_observation_2009} in the low-density regime (solid line). \textbf{c}, \textbf{d} Distributions of mobility $\mu$ measured at $n = 6 \times 10^{11}$ cm$^{-2}$ and percolation density $n_\textnormal{p}$ for heterostructures featuring quantum wells of different thickness $w$: 9.0(5)~nm (blue, 16 H-FETs measured and reported in ref.~\cite{paquelet_wuetz_reducing_2023}), 6.9(5)~nm (purple, from the analysis of the same dataset in \textbf{a}, \textbf{b}), and 5.3(5)~nm (green, 22 H-FETs measured and reported in ref.~\cite{paquelet_wuetz_reducing_2023}). 
Violin plots, quartile box plots, and mode (horizontal line) are shown.}
\end{figure}

\newpage
\begin{figure}[t]
	\includegraphics[width=86mm]{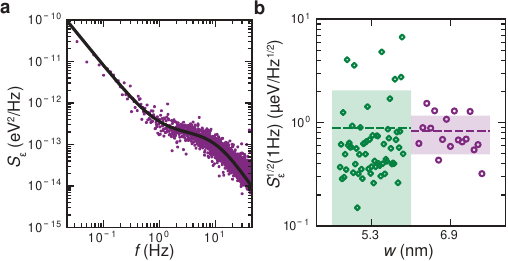}%
\caption{\textbf{Charge noise measurements.} \textbf{a} Charge noise power spectral density $S_{\epsilon}$ measured on a flank of a Coulomb peak and extracted using the lever arm of the corresponding Coulomb diamond. The black line is a fit to the function, which is the sum of a power law and a Lorentzian. 
\textbf{b} Experimental scatter plots of charge noise at $1$ Hz ($S_{\epsilon}^{1/2} (\textnormal{1 Hz})$) obtained by repeating charge noise spectrum measurements as in \textbf{a} for multiple devices and different electron occupancy. Data from the 6.9(5)~nm quantum well (purple, 2 devices, 17 spectra) is compared to data from the 5.3(5)~nm quantum well (green, 63 spectra, 5 devices, reported in ref.~\cite{paquelet_wuetz_reducing_2023}). 
We compare single-layer devices (diamonds) and multi-layer devices featuring overlapping gate geometry and micromagnets (circles). Dashed lines and shaded area denote the mean value and two standard deviations.}
\end{figure}

\newpage
\begin{figure}[t]
	\includegraphics[width=86mm]{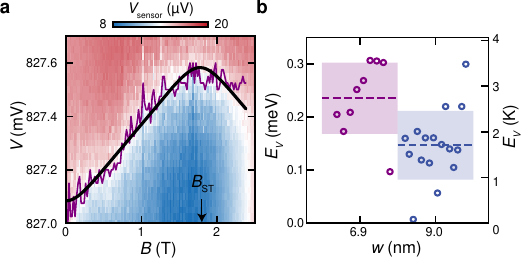}%
\caption{\textbf{Valley splitting measurements.} \textbf{a} Typical magnetospectroscopy map of the (1e) $\rightarrow$ (2e) charge transition, used to measure singlet-triplet splittings. 
The thin purple line shows the location of the charge transition at a fixed magnetic field. 
The thick black line is a fit to the data (Methods), from which we extract the kink position $B_\textnormal{{ST}}$. 
The valley splitting $E_\textnormal{{V}}$ is given by $E_\textnormal{{V}} = g\mu_\textnormal{B} B_\textnormal{{ST}}$, where $g = 2$ is the gyromagnetic ratio, and $\mu_\textnormal{B}$ is the Bohr magneton. 
\textbf{b} Experimental scatter plots of valley splitting obtained by magnetospectroscopy on complete spin qubit devices. Data from heterostructures with a 6.9(5)~nm quantum well (purple, 9 quantum dots from 2 devices) is compared to data from a 9.0(5)~nm quantum well (blue, 16 quantum dots, 3 devices, from ref.~\cite{philips_universal_2022}). Dashed lines and shaded area denote the mean value and two standard deviations.}
\end{figure}

\end{document}